\documentclass[11pt]{article}
\usepackage{moriond,epsfig}

\def\Journal#1#2#3#4{{#1} {\bf #2}, #3 (#4)}

\def\NPB{{\em Nucl. Phys.} B}
\def\PLB{{\em Phys. Lett.}  B}
\def\PRL{\em Phys. Rev. Lett.}
\def\PRD{{\em Phys. Rev.} D}
\def\ZPC{{\em Z. Phys.} C}

\def\be{\begin{equation}}
\def\ee{\end{equation}}
\def\bea{\begin{eqnarray}}
\def\eea{\end{eqnarray}}

\def\elel{\ell^+\ell^-}

\def\KP{K^+}
\def\piP{\pi^+}
\def\piM{\pi^-}
\def\piZ{\pi^0}

\def\KS{K^0_S}

\def\Kst{K^*}
\def\KstZ{K^{*0}}
\def\KstP{K^{*+}}

\def\Kstll{\Kst\elel}

\def\fbinv{{\rm~fb}^{-1}}

\def\Mbc{M_{\rm bc}}

\def\Cseven{\tilde{C}_7^{\rm eff}}
\def\Cnine{\tilde{C}_9^{\rm eff}}
\def\Cten{\tilde{C}_{10}^{\rm eff}}
\def\Ci{\tilde{C}_i}

\def\PM#1#2{\,^{+#1}_{-#2}}

\def\Journal#1#2#3#4{{#1} {\bf #2}, #3 (#4)} % {journal}{vol}{page}{year}
\def\NPB{Nucl. Phys. B}
\def\PLB{Phys. Lett. B}
\def\PRL{Phys. Rev. Lett.}
\def\PRD{Phys. Rev. D}
\def\ZPC{Z. Phys. C}

\def\RMP{Rev. Mod. Phys.}

\def\be{\begin{equation}}
\def\ee{\end{equation}}
\def\ba{\begin{eqnarray}}
\def\ea{\end{eqnarray}}

\begin{document}
\vspace*{4cm}
\title{MEASUREMENT OF FORWARD-BACKWARD ASYMMETRY IN \boldmath$B\to K^*\ell^+\ell^-$ AND 
EVIDENCE OF \boldmath$B^{-}\rightarrow\tau^{-}\bar{\nu}_{\tau}$}

\author{ K. IKADO}

\address{Department of Physics, Nagoya University, \\
Nagoya, 464-8602, Japan}

\maketitle\abstracts{
We report the first measurement of the forward-backward asymmetry and 
the ratios of Wilson coefficients
%$A_9/A_7$ and $A_{10}/A_7$ 
in $B \to \Kstll$, 
%where $\ell$ represents an electron or a muon
using 386
$\times 10^6$ $B\bar{B}$ pairs that were collected on the $\Upsilon(4S)$ resonance
with the Belle detector at the KEKB asymmetric-energy $e^+ e^-$ collider.
%We observe a large integrated forward-backward asymmetry with a significance of
%3.4$\sigma$.
%The results are obtained from a data sample containing 386
%$\times 10^6$ $B\bar{B}$ pairs that were collected on the $\Upsilon(4S)$ resonance
%with the Belle detector at the KEKB asymmetric-energy $e^+ e^-$ collider.
We also present the first evidence of the decay $B^{-}\rightarrow\tau^{-}\bar{\nu}_{\tau}$, 
using $414~\textrm{fb}^{-1}$ of data.
% collected with the Belle detector.
%Events are tagged by fully reconstructing one of the $B$ mesons in hadronic modes.
%We detect the signal with a significance of $4.0$$\sigma$ including systematics, 
%and measure the branching fraction to be 
%${\cal B}(B^{-}\rightarrow\tau^{-}\bar{\nu}_{\tau}) = 
% (1.06^{+0.34}_{-0.28}(\mbox{stat})^{+0.22}_{-0.25}(\mbox{syst}))\times 10^{-4}$.
}

\section{Introduction}
Flavor-changing neutral current $b \to s$ processes proceed 
via loop diagrams in the Standard Model (SM). 
The $b \to s$ processes are sensitive to new physics effect. If new heavy particles can
contribute to the decays, their amplitudes will interfere with the SM amplitudes
and thereby modify the decay rate as well as decay distributions.
Such contributions may change the
Wilson coefficients~\cite{BBL} that parametrize the strength of the short
distance interactions. The $b \to s \ell^+ \ell^-$ amplitude is described
by the effective Wilson coefficients $\Cseven$, $\Cnine$ and 
$\Cten$.
%, whose terms have been calculated up to next-to-next-to-leading order (NNLO)~\cite{WC} in the SM. 
%The magnitude of $\Cseven$ is strongly constrained from measurements of 
%$B \to X_{s}\gamma$~\cite{btosgamma,C7}
%and a large area of the ($\Cnine$, $\Cten$) plane is excluded by branching fraction
%measurements of $B\to K^{(*)} \ell^+\ell^-$
%and $B \to \Xsll$~\cite{BelleKstarll,KllXsll,ALGH,Gambino}.
%However the sign of $\Cseven$ and values of $\Cnine$ and $\Cten$ are not yet determined.
Measurement of the forward-backward asymmetry and differential decay rate as
 functions of $q^2$ and $\theta$ for $B \to \Kstll$ constrains the relative signs and magnitudes of these coefficients~\cite{C10flip,C7C9C10}. 
Here, $q^2$ is the squared invariant mass of  the dilepton system, 
and $\theta$ is the angle between the momenta of the negative (positive)
lepton and the $B$ ($\bar{B}$) meson in the dilepton rest frame.
%The forward-backward asymmetry is defined using the differential decay width, $g(q^2,\theta)= d^2\Gamma/dq^2d\cos\theta$~\cite{bib:abhh}, as 
%\begin{eqnarray}
%{\cal A}_{\rm FB}(q^2) = { \int_{-1}^1 {\rm sgn}(\cos\theta) g(q^2,\theta)\,d\cos\theta \over \int_{-1}^1 g(q^2,\theta)\,d\cos\theta}.
%\label{eq:afb}
%\end{eqnarray}
%The numerator in Eq.~\ref{eq:afb} does not cancel due to interference between the electroweak penguin and box diagrams, and can be expressed in terms of Wilson coefficients as
%\begin{eqnarray}
%& &  \int^1_{-1} {\rm sgn}(\cos\theta) g(q^2,\theta) d\cos\theta \nonumber \\
%&=&  -\Cten \xi(q^2) \left( {\rm Re}(\Cnine)F_1 + {1\over q^2}\Cseven F_2 \right) ,
%\end{eqnarray}
%where $\xi$ is a function of $q^2$, and $F_{1,2}$ are functions of form factors (the full expression can be found in Ref.~\cite{bib:abhh}).

The purely leptonic decay 
$B^{-}\rightarrow\tau^{-}\bar{\nu}_{\tau}$
proceeds via annihilation of $b$ and $\overline{u}$ quarks to a $W^-$ boson in the SM.
It provides a direct determination of the product of the $B$ meson decay 
constant $f_B$ and the magnitude of the
Cabibbo-Kobayashi-Maskawa (CKM) matrix element $|V_{ub}|$.
The branching fraction is given by
\begin{equation}
 \label{eq:BR_B_taunu}
{\cal B}(B^{-}\rightarrow\tau^{-}\bar{\nu}_{\tau}) 
= \frac{G_{F}^{2}m_{B}m_{\tau}^{2}}{8\pi}\left(1-\frac{m_{\tau}^{2}}
{m_{B}^{2}}\right)^{2}f_{B}^{2}|V_{ub}|^{2}\tau_{B},
\end{equation}
where $G_F$ is the Fermi coupling constant, 
$m_{B}$ and $m_{\tau}$ are the $B$ and $\tau$ masses, respectively, 
and $\tau_B$ is the $B^-$ lifetime~\cite{Eidelman:2004wy}.
%The expected branching fraction is  $(1.59\pm 0.40) \times 10^{-4}$
%using $|V_{ub}| = (4.39 \pm 0.33) \times 10^{-3}$, determined by inclusive
%charmless semileptonic $B$ decay data~\cite{HFAG}, $\tau_{B} = 1.643\pm 0.010$ ps ~\cite{HFAG},
%and $f_B = 0.216\pm 0.022$ GeV
%obtained from lattice QCD calculations~\cite{Gray:2005ad}.
%Physics beyond the SM, such as supersymmetry or two-Higgs doublet models,
%could modify ${\cal B}(B^{-}\rightarrow\tau^{-}\bar{\nu}_{\tau})$ through
%the introduction of a charged Higgs boson~\cite{Hou:1992sy}.
Purely leptonic $B$ decays have not been observed in past experiments.
The most stringent upper limit on $B^{-}\rightarrow\tau^{-}\bar{\nu}_{\tau}$
comes from the BaBar experiment: 
${\cal B}(B^{-}\rightarrow\tau^{-}\bar{\nu}_{\tau}) < 2.6 \times 10^{-4}$
(90\% C.L.)~\cite{Aubert:2005}.

The Belle detector is a large-solid-angle
magnetic spectrometer consisting of a silicon vertex detector,
a $50$-layer central drift chamber (CDC), a system of aerogel threshold
$\check{\textrm{C}}$erenkov counters (ACC), time-of-flight scintillation
counters (TOF), and an electromagnetic calorimeter comprised of
CsI(Tl) crystals (ECL)
located inside a superconducting solenoid coil that provides a $1.5$ T
magnetic field. An iron flux-return located outside of the coil is
instrumented to identify $K_{L}^{0}$ and muons.
The detector is described in detail elsewhere~\cite{belle_detector:2003}.

\section{Measurement of Forward-Backward Asymmetry in \boldmath$B\to K^*\ell^+\ell^-$}

                      %%%%%%%%%%%%%%%%%%%%%%%%
                      %%%  RECONSTRUCTION  %%%
                      %%%%%%%%%%%%%%%%%%%%%%%%

We use a $357\fbinv$ data sample containing 386 $\times 10^6$ 
$B\bar{B}$ pairs taken at the $\Upsilon(4S)$ resonance.
We also study the $B^+ \to K^+ \ell^+ \ell^-$ mode, which
is expected to have very small forward-backward asymmetry even in the existence of
new physics~\cite{NoAFBKll}.
%The event reconstruction procedure is the same as described in our previous report~\cite{BelleKstarll}.  
The following final states are used to reconstruct $B$ candidates:
$\KstZ \elel$, $\KstP \elel$, and $\KP \elel$,
with subdecays
$\KstZ\to\KP\piM$, $\KstP\to\KS\piP$ and $\KP\piZ$,
$\KS\to\piP\piM$, and $\piZ\to\gamma\gamma$.
Hereafter, $\KstZ \elel$ and $\KstP \elel$ are combined and called
$K^* \elel$.
We use two variables defined in the center-of-mass (CM) frame to select $B$
candidates: the beam-energy constrained mass $M_{\rm bc}\equiv\sqrt{E_{\rm beam}^{2} - p_{B}^{2}}$
and the energy difference $\Delta E\equiv E_{B} - E_{\rm beam}$, where 
$p_{B}$ and $E_B$ are the measured CM momentum and energy of the $B$
candidate, and $E_{\rm{beam}}$ is the CM beam energy. 
%When multiple candidates are
%found in an event, we select the candidate with the smallest value of $|\Delta E|$.
                        %%%%%%%%%%%%%%%%%%%%
                        %%%  BACKGROUND  %%%
                        %%%%%%%%%%%%%%%%%%%%
The dominant background consists of $B\bar{B}$ events where both $B$ mesons decay semileptonically. 
We suppress this using missing energy and $\cos\theta_B^*$, where $\theta_B^*$ is the angle between 
the flight direction of the $B$ meson and the beam axis in the CM frame.
%These quantities are combined to form signal and background likelihoods, ${\cal{L}}_{\rm{sig}}$ and
%${\cal{L}}_{B\bar{B}}$, and
%event selection is then performed using the ratio ${\cal{R}}_{B\bar{B}} =
%{\cal{L}}_{\rm{sig}} / ({\cal{L}}_{\rm{sig}} + {\cal{L}}_{B\bar{B}})$.
%The continuum ($e^+e^- \to q\bar{q}$, $q=u,d,s,c$) background is suppressed using
%a likelihood ratio ${\cal{R}}_{\rm cont}$ 
%(defined similarly to ${\cal{R}}_{B\bar{B}}$)
%that depends on three variables;
%a Fisher discriminant~\cite{fd} calculated from the energy flow in 9 cones along the $B$ candidate sphericity axis and 
%the normalized second Fox-Wolfram moment~\cite{fw},
%the angle between the beam axis
%and the CM sphericity axis calculated with tracks used in the $B$ 
%meson reconstruction, and $\cos\theta_B^*$.
Backgrounds from $B \to J/\psi X_s, \psi(2S) X_s$ decays are rejected using the
dilepton invariant mass.
%Backgrounds from photon conversions and $\pi^{0}$ Dalitz
%decays are suppressed by requiring the $e^+e^-$ invariant mass to be above 140~MeV/$c^2$. 
The signal box is defined as $|M_{\rm bc}-m_{B}|<8$~MeV/$c^{2}$ for
both lepton modes and
$-55\, (-35)\, {\rm MeV} < \Delta E < 35\, {\rm MeV}$ for the electron (muon) mode.
%We optimize the selections on ${\cal{R}}_{\rm cont}$ and ${\cal{R}}_{B\bar{B}}$
%for each $K^*$ decay mode and each lepton mode to maximize sensitivity to events with $q^2 <$  6~GeV${}^2$/$c^2$.

                         %%%%%%%%%%%%%%%%%
                         %%%  FITTING  %%%
                         %%%%%%%%%%%%%%%%%

\begin{figure}[htbp]
\begin{center}
  \includegraphics[scale=0.4]{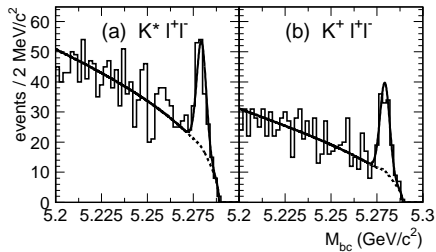}
\end{center}
\caption{$M_{\rm bc}$ distributions for (a) $B \to K^{*} \ell^{+} \ell^{-}$ 
and (b) $B\to K^+ \ell^{+} \ell^{-}$ samples. The solid
and dashed curves are the fit results for the total and background contributions.}
\label{fig:mbcfit}
\end{figure}

We perform an unbinned maximum-likelihood fit to the $M_{\mathrm{bc}}$ distribution to determine the signal yield.
The fit function includes signal, cross-feeds and other background components. 
%The cross-feeds are misreconstructed $\KorKstarll$ events with correct (``CF'') and incorrect (``IF'') flavor tagging. 
%The cross-feed from $\Xsll$ events other than $\KorKstarll$ is negligible. 
%The other backgrounds come from dilepton background, combinatorial $K^{(*)} \ell^{\pm} h^{\mp}$, $K^{(*)} h^{+} h^{-}$ 
%and $\psi X_s$ events, where $h$ represents a pion or a kaon. 
%The dilepton background refers to the sum of all background sources with two leptons where the lepton is 
%from (semi)leptonic meson decays, photon conversions and $\pi^0$ Dalitz decays. 
%The $K^{(*)} h^{+} h^{-}$ is from both combinatorial background and $B$ meson decays.
%The shape for cross-feed events is parametrized by a sum of an ARGUS function~\cite{ARGUS} and a Gaussian 
%whose parameters are determined from Monte Carlo (MC) samples.
%The dilepton background is characterized by an ARGUS function.
%The shape of each background is determined from a MC sample.
%% (The $K^{(*)} e^{\pm} \mu^{\mp}$ background shape is found to be consistent in MC and data.)
%Since the shape for $K^{(*)} \ell^{\pm} h^{\mp}$ is similar to that for the dilepton background, 
%we use the same parameterizations for both backgrounds. 
%The residual background from $\psi X_s$ is estimated from a MC sample of
%$\psi$ inclusive events and parametrized by the sum of an ARGUS
%function and a Gaussian. 
%The background from events with misidentified leptons is also parametrized by the sum of an ARGUS function and a Gaussian.
In the fit, all background fractions except the dilepton background are fixed while the signal fraction is allowed to float.
                          %%%%%%%%%%%%%%%
                          %%%  YIELD  %%%
                          %%%%%%%%%%%%%%%
We obtain $113.6 \pm 13.0$ and $96.0 \pm 12.0$ signal events for $\Kstll$ and $K^+\ell^+\ell^-$, respectively.
Figure~\ref{fig:mbcfit} shows the fit result. 

                       %%%%%%%%%%%%%%%%%%%%%
                       %%%  FITTOWIDTH   %%%
                       %%%%%%%%%%%%%%%%%%%%%

We use $B \to K^* \ell^+ \ell^-$ candidates in the signal box to measure the normalized double differential decay width. 
For the evaluation of the Wilson coefficients, the NNLO Wilson coefficients 
$\Ci$ of Ref.~\cite{WC} are used. 
Since the full NNLO calculation only exists for $q^2/m_b^2<0.25$ region, we adopt the so-called partial 
NNLO calculation~\cite{ALGH} for $q^2/m_b^2>0.25$.
The higher order terms in the $\Ci$ are fixed to the SM values while the leading terms $A_i$, 
with the exception of $A_7$, are allowed to float. Since the branching fraction measurement of 
$B \to X_s \gamma$ is consistent with the prediction within the SM, $A_7$ is fixed at the SM value, 
$-0.330$, or the sign-flipped value, $+0.330$. 
We choose $A_9/A_7$ and $A_{10}/A_7$ as fit parameters.
The SM values for $A_9$ and $A_{10}$ are 4.069 and -4.213, respectively~\cite{ALGH}.
To extract the these ratios, we perform an unbinned maximum likelihood fit to the events 
in the signal box with a probability density function (PDF) that includes the normalized double differential decay width.

\begin{figure}[htbp]
\begin{center}
  \includegraphics[scale=0.55]{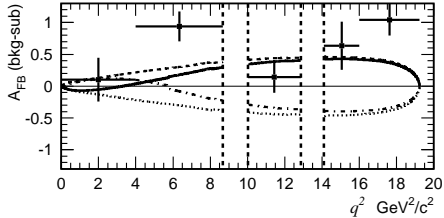}
\end{center}
\caption{Fit result for the negative $A_7$ solution (solid) projected onto the background subtracted forward-backward asymmetry, 
  and forward-backward asymmetry curves for several input parameters, including the effects of efficiency; $A_7$ positive 
  case~($A_7=0.330$, $A_9=4.069$, $A_{10}=-4.213$) (dashed), $A_{10}$ positive case ($A_7=-0.280$, $A_9=2.419$, $A_{10}=1.317$) 
  (dot-dashed) and both $A_7$ and $A_{10}$ positive case~($A_7=0.280$, $A_9=2.219$, $A_{10}=3.817$) (dotted). 
  The new physics scenarios shown by the dot-dashed and dotted curves are excluded.}
\label{fig:afb}
\end{figure}

We measure the integrated asymmetry $\tilde{{\cal A}}_{\rm FB}$, which is defined as
\begin{eqnarray}
\tilde{{\cal A}}_{\rm FB} ={\int  \int^1_{-1} {\rm sgn}(\cos\theta)g(q^2,\theta) d\cos\theta dq^2\over \int \int^1_{-1} g(q^2,\theta) d\cos\theta dq^2}.
\end{eqnarray}
We determine the yield in each $q^2$ and forward-backward regions from a fit to the $\Mbc$ distribution. 
Then we correct the efficiency and obtain
\begin{eqnarray}
\tilde{{\cal A}}_{\rm FB}(B \to \Kstll) &=& 0.50 \pm 0.15 \pm 0.02, \nonumber \\
\tilde{{\cal A}}_{\rm FB}(B^+ \to K^+ \ell^+ \ell^-) &=& 0.10 \pm 0.14 \pm 0.01,
\end{eqnarray}
where the first error is statistical and the second is systematic.
A large integrated asymmetry is observed for $\Kstll$ with a significance of $3.4\sigma$.
The result for $K^+ \ell^+ \ell^-$ is consistent with zero as expected. 
%We fit the $\Kstll$ candidates with the PDF of Eq.~\ref{eq:pdf}.
The fit results of ratios of Wilson coefficients are summarized in Table~\ref{tab:result}.
Figure~\ref{fig:afb} shows the fit results projected onto the background-subtracted forward-backward 
asymmetry distribution in bins of $q^2$. 
\begin{table}[htbp]
  \caption{%
    $A_9/A_7$ and $A_{10}/A_7$ fit results for negative and positive $A_7$ values. 
    The first error is statistical and the second is systematic.}
 \begin{center}
 \begin{tabular}{ccccc} \hline 
		 & & Negative $A_7$       & & Positive $A_7$ \\ \hline
$A_9/A_7$        &  & $-15.3 \PM{3.4}{4.8}\pm1.1$ &  & $-16.3 \PM{3.7}{5.7}\pm1.4$ \\ 
$A_{10}/A_7$     &  & $10.3 \PM{5.2}{3.5}\pm1.8$  &  & $11.1 \PM{6.0}{3.9}\pm2.4$ \\ \hline 
  \end{tabular}
   \end{center}
\label{tab:result}
\end{table}

\begin{figure}[htbp]
\begin{center}
  \includegraphics[scale=0.2]{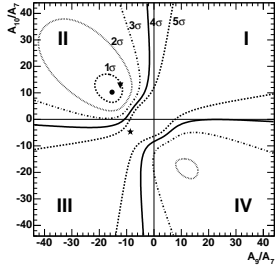}
\end{center}
\caption{Confidence level contours for negative $A_7$. Curves show 1$\sigma$ to 5 $\sigma$ contours. 
  The symbols show the fit (circle), SM (triangle), and $A_{10}$-positive (star) cases.}
\label{fig:cl}
\end{figure}

The fit results are 
consistent with the SM values ${A_9}/{A_7}=-12.3$ and ${A_{10}}/{A_7}=12.8$.
In Fig.~\ref{fig:cl}, we show confidence level (CL) contours in the ($A_9/A_7$, $A_{10}/A_7$) plane 
based on the fit likelihood smeared by the systematic error, which is assumed to have a Gaussian distribution.
We also calculate an interval in $A_9A_{10}/A_7^2$ at the 95\% CL for the allowed $A_7$ region,
\begin{eqnarray}
-14.0 \times 10^2 < {A_9}{A_{10}}/{A_7^2} < -26.4.
\end{eqnarray}
From this, the sign of ${A_9}{A_{10}}$ must be negative, and the solutions in quadrants I and III 
of Fig.~\ref{fig:cl} are excluded at 98.2\% confidence level.
Since solutions in both quadrants II and IV are allowed, we cannot determine the sign of $A_7A_{10}$.
Figure~\ref{fig:afb} shows the comparison between the fit results for the negative $A_7$ value projected 
onto the forward-backward asymmetry, and the forward-backward asymmetry distributions for several input parameters. 
We exclude the new physics scenarios shown by the dotted and dot-dashed curves, which have a positive ${A_9}{A_{10}}$ value.

                          %%%%%%%%%%%%%%%%%
                          %%%  SUMMARY  %%%
                          %%%%%%%%%%%%%%%%%

%In summary, we have measured the ratios of Wilson coefficients in $B \to \Kstll$ decay for the first time by studying the forward-backward asymmetry in the angular distribution of leptons. We observe a large integrated forward-backward asymmetry with a significance of $3.4\sigma$.
%The fit results are consistent with the SM prediction and also with the case where the sign of $A_7A_{10}$ is flipped. We exclude new physics scenarios with positive $A_9A_{10}$ at 98.2\% confidence.

\section{Evidence of the Purely Leptonic Decay \boldmath$B^{-}\rightarrow\tau^{-}\bar{\nu}_{\tau}$}

%===========================================================================
We use a $414~\textrm{fb}^{-1}$ data sample containing 
$447\times 10^{6}$ $B$ meson pairs collected with the Belle detector.
%at the KEKB asymmetric-energy $e^{+}e^{-}$ ($3.5$ on $8$ GeV) collider
%operating at the $\Upsilon(4S)$ resonance ($\sqrt{s} = 10.58$ GeV). 
We use a detailed MC simulation, which fully describes the
detector geometry and response based on GEANT~\cite{GEANT}, to determine
the signal selection efficiency and study the background.
%In order to reproduce effects of beam background, data taken with random
%triggers for each run period are overlaid on simulated events. 
The $B^{-}\rightarrow\tau^{-}\bar{\nu}_{\tau}$ signal decay is generated 
by the EvtGen package~\cite{EvtGen}.
To model the background from $e^+e^- \to B\overline{B}$ and continuum 
$q\overline{q}~(q = u, d, s, c)$ production processes, large 
$B\overline{B}$ and $q\overline{q}$ MC samples 
corresponding to about twice the data sample
are used.
%We also use MC samples for rare $B$ decay processes, such as charmless 
%hadronic, radiative, electroweak decays and $b \to u$ semileptonic decays.

We fully reconstruct one of the $B$ mesons in 
the event, referred to hereafter as the tag side ($B_{\rm tag}$), 
and compare properties of the remaining particle(s), referred to as the 
signal side ($B_{\rm sig}$), to those expected for signal and background.
In the events where a $B_{\rm tag}$ is reconstructed, we search for decays
of $B_{\rm sig}$ into a $\tau$ and a neutrino. 
Candidate events are required to have one or three charged track(s) on the
signal side with the total charge being opposite to that of $B_{\rm tag}$.
The $\tau$ lepton is identified in the five decay modes,
$\mu^{-}\bar{\nu}_{\mu}\nu_{\tau}$,
$e^{-}\bar{\nu}_{e}\nu_{\tau}$, 
$\pi^{-}\nu_{\tau}$,
$\pi^{-}\pi^{0}\nu_{\tau}$ and 
$\pi^{-}\pi^{+}\pi^{-}\nu_{\tau}$,
which taken together correspond to $81\%$ of all $\tau$ decays~\cite{Eidelman:2004wy}.
The muon, electron and charged pion candidates are selected based on
information from particle identification devices.
%The leptons are selected with requirements that have efficiencies greater than 90\% 
%for both muons and electrons in the momentum region above 1.2 GeV/$c$, and 
%misidentification rates of less than 0.2\%(1.5\%) for electrons (muons) in 
%the same momentum region. 
%Kaon candidates are rejected for all charged tracks on the signal side.
%The $\pi^0$ candidates are reconstructed by requiring the invariant mass of two
%$\gamma$'s to satisfy $|M_{\gamma\gamma}-m_{\pi^0}| < 20~\mbox{MeV}/c^{2}$.
For all modes except $\tau^{-}\rightarrow\pi^{-}\pi^{0}\nu_{\tau}$, we reject events with 
$\pi^{0}$ mesons on the signal side.
%We place the following requirements on the track momentum in the CM frame,
%$p_{\ell} > 0.3~\textrm{GeV}/c$ for $\mu^{-}\bar{\nu}_{\mu}\nu_{\tau}$ and $e^{-}\bar{\nu}_{e}\nu_{\tau}$,
%$p_{\pi^{-}} > 1.0~\textrm{GeV}/c$ for $\pi^{-}\nu_{\tau}$,
%$p_{\pi^{-}\pi^{0}} > 1.2~\textrm{GeV}/c$ for $\pi^{-}\pi^{0}\nu_{\tau}$ and
%$p_{\pi^{-}\pi^{+}\pi^{-}} > 1.8~\textrm{GeV}/c$ for
%$\pi^{-}\pi^{+}\pi^{-}\nu_{\tau}$.
%We calculate the missing momentum of the event in the CM frame ($p_{\rm miss}$)
%from $p_B$ and the momenta of charged tracks and $\pi^0$'s on the signal side.
%We require
%$p_{\rm miss} > 0.2~\textrm{GeV}/c$ for $\mu^{-}\bar{\nu}_{\mu}\nu_{\tau}$ and 
%$e^{-}\bar{\nu}_{e}\nu_{\tau}$,
%$p_{\rm miss} > 1.0~\textrm{GeV}/c$ for $\pi^{-}\nu_{\tau}$,
%$p_{\rm miss} > 1.2~\textrm{GeV}/c$ for $\pi^{-}\pi^{0}\nu_{\tau}$ and
%$p_{\rm miss} > 1.8~\textrm{GeV}/c$ for $\pi^{-}\pi^{+}\pi^{-}\nu_{\tau}$.
%In order to suppress background where particles produced along the beam pipe
%escape detection, the cosine of the angle of the missing momentum 
%($\cos\theta_{\rm miss}^{*}$) is required to satisfy
%$-0.86 < \cos\theta_{\rm miss}^{*} < 0.95$
%in the CM frame.
%We further require the invariant mass of the visible decay products to satisfy
%$|M_{\pi\pi}-m_{\rho}| < 0.15~\mbox{GeV}/c^{2}$ and 
%$|M_{\pi\pi\pi}-m_{a_{1}^{-}}| < 0.3~\mbox{GeV}/c^{2}$.
%All the selection criteria have been optimized to achieve the highest sensitivity in MC.

\begin{figure}[htbp]
\begin{center}
  \includegraphics[scale=0.22]{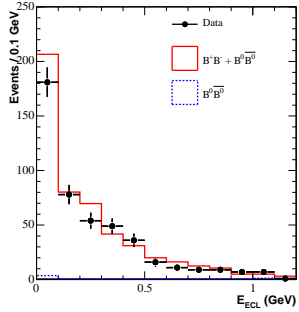}
\caption{$E_{\rm ECL}$ distribution for the both $B$ tagged events, where one $B$ is fully
  reconstructed and the other $B$ is reconstructed as $B^{-} \rightarrow D^{*0}\ell^{-}\bar{\nu}$. 
  The dots with errors indicate the data. The solid histogram represents the background from 
  $B\overline{B}$ MC ($B^+B^- + B^0\overline{B}{}^0$), and the dashed histogram shows the contribution 
  from $B^0\overline{B}{}^0$ events.}
    \label{fig:controlsample}
\end{center}   
\end{figure}

The most powerful variable for separating signal and background is the 
remaining energy in the ECL, denoted as $E_{\rm ECL}$, which is sum of
the energy of photons that are not associated with either the 
$B_{\rm tag}$ or the $\pi^{0}$ candidate from the 
$\tau^{-}\rightarrow \pi^{-}\pi^{0}\nu_{\tau}$ decay.
%For neutral clusters contributing to $E_{\rm ECL}$, we require a minimum 
%energy threshold of 50 MeV for the barrel and 100 (150) MeV for the forward (backward) endcap ECL.
%A higher threshold is used for the endcap ECL because the effect of beam
%background is more severe.
For signal events, $E_{\rm ECL}$ must be either zero or a small value 
arising from beam background hits, therefore, signal events peak at 
low $E_{\rm ECL}$.
On the other hand background events are distributed toward higher 
$E_{\rm ECL}$ due to the contribution from additional neutral clusters.
%===========================================================================
The $E_{\rm ECL}$ signal region is optimized for each $\tau$ decay mode based
on the MC simulation, and is defined by $E_{\rm ECL} < 0.2~\mbox{GeV}$ for the
$\mu^{-}\bar{\nu}_{\mu}\nu_{\tau}$, $e^{-}\bar{\nu}_{e}\nu_{\tau}$ and
$\pi^{-}\nu_{\tau}$ modes, and $E_{\rm ECL} < 0.3~\mbox{GeV}$ for
the $\pi^{-}\pi^{0}\nu_{\tau}$ and $\pi^{-}\pi^{+}\pi^{-}\nu_{\tau}$
modes. 
The $E_{\rm ECL}$ sideband region is defined by $0.4~\mbox{GeV} < E_{\rm ECL} < 1.2$ GeV 
for the $\mu^{-}\bar{\nu}_{\mu}\nu_{\tau}$, $e^{-}\bar{\nu}_{e}\nu_{\tau}$ and
$\pi^{-}\nu_{\tau}$ modes, and by $0.45~\mbox{GeV} < E_{\rm ECL} < 1.2$ GeV for
the $\pi^{-}\pi^{0}\nu_{\tau}$ and $\pi^{-}\pi^{+}\pi^{-}\nu_{\tau}$ modes.
Table~\ref{tab:signal_yields} shows the number of events found in the sideband 
region for data ($N_{\rm side}^{\rm obs}$) and for the background MC simulation 
($N_{\rm side}^{\rm MC}$).
% scaled to the equivalent integrated luminosity in data.
%Their good agreement for each $\tau$ decay mode indicates the validity of the
%background MC simulation.
Table~\ref{tab:signal_yields} also shows the number of the background MC events 
in the signal region ($N_{\rm sig}^{\rm MC}$).  
%According to the MC simulation, about 95\% (5\%) of the background events come 
%from $B\overline{B} (q\overline{q})$ processes. 
%The MC simulation predicts that the background in the signal region consists of
%$B^- \to D^{(*)0} \ell^- \bar{\nu}$ semileptonic decays (90\%) and rare $B$ decay
%processes (10\%). 
%About 30\% of the background have $K_L^0$ candidates in the KLM.
%==========================================================================
In order to validate the $E_{\rm ECL}$ simulation, we use a control sample
of events (double tagged events), where the $B_{\rm tag}$ is fully reconstructed 
as described above and $B_{\rm sig}$ is reconstructed in the decay chain, 
$B^{-} \rightarrow D^{*0}\ell^{-}\bar{\nu}$ ($D^{*0}\rightarrow D^{0}\pi^{0}$),
followed by $D^0 \to K^- \pi^+$ or $K^- \pi^- \pi^+ \pi^+$
where $\ell$ is a muon or electron.
%The sources affecting the $E_{\rm ECL}$ distribution in the control sample 
%are similar to those affecting the $E_{\rm ECL}$ distribution in the signal 
%MC simulation.
Figure~\ref{fig:controlsample} shows the $E_{\rm ECL}$ distribution in the
control sample for data and the MC simulation scaled to 
equivalent integrated luminosity in data.
%Their agreement demonstrates the validity of the $E_{\rm ECL}$ simulation 
%in the signal MC.

% --------------------------------------------------

\begin{figure}[htbp]
\begin{center}
  \includegraphics[scale=0.2]{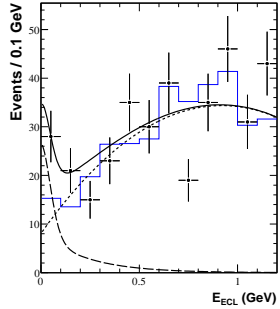}
\caption{$E_{\rm ECL}$ distributions in the data after
	all selection requirements 
	%except the one on $E_{\rm ECL}$ 
	have been applied. The data and background MC samples are represented by the points with error bars
	and the solid histogram, respectively. 
	%The dashed histogram indicates 
	%the $B^{-}\rightarrow\tau^{-}\overline{\nu}_{\tau}$ signal distribution from MC. 
	The solid curve shows the result of the fit with the sum of the signal shape (dashed) and background shape
	(dotted).
}
    \label{ecl_opened}
\end{center}   
\end{figure}

After finalizing the signal selection criteria, the signal region is examined.
Figure~\ref{ecl_opened} shows the obtained $E_{\rm ECL}$ distribution
when all $\tau$ decay modes are combined.
One can see a significant excess of events in the $E_{\rm ECL}$ signal region
below $E_{\rm ECL}< 0.25$ GeV.
Table~\ref{tab:signal_yields} shows the number of events observed in the 
signal region ($N_{\rm obs}$) for each $\tau$ decay mode.
%For the events in the signal region, we verify that the distributions of the 
%event selection variables other than $E_{\rm ECL}$, such as $M_{\rm bc}$ and
%$p_{\rm miss}$, are consistent with the sum of the signal and background
%distributions expected from MC.
%The excess is robust against a cut requiring no $K_L^0$ candidate.

% -------------------------------------------------

%==========================================================================
We deduce the final results by fitting the obtained $E_{\rm ECL}$ 
distributions to the sum of the expected signal and background shapes.
Probability density functions (PDFs) for the signal $f_s(E_{\rm ECL})$ and
for the background $f_b(E_{\rm ECL})$ are constructed for each $\tau$ decay
mode from the MC simulation.
The signal PDF is modeled as the sum of a Gaussian function, centered at
$E_{\rm ECL} = 0$, and an exponential function.
The background PDF, as determined from the MC simulation, is parametrized 
by a second-order polynomial.
%The PDFs are combined into an extended likelihood function,
%\begin{equation}
%{\cal L} = \frac{e^{-(n_{s}+n_{b})}}{N!}\prod_{i=1}^{N}(n_{s}f_{s}(E_{i})+n_{b}f_{b}(E_{i})),
%\end{equation}
%where $E_{i}$ is the $E_{\rm ECL}$ in the $i$th event, $N$ is the total number 
%of events in the data, and $n_{s}$ and $n_{b}$ are the signal yield and background 
%yield to be determined by the fit.
%To combine likelihood functions of the five decay modes, we multiply 
%the likelihood functions to produce the combined likelihood 
%(${\cal L}_{\rm com} = \prod_{j=1}^{5} {\cal L}_{j}$).
The results are listed in Table~\ref{tab:signal_yields}.
The number of signal events in the signal region deduced from the fit ($N_{\rm s}$) is $21.2^{+6.7}_{-5.7}$
when all $\tau$ decay modes are combined.
Table~\ref{tab:signal_yields} also gives the number of background events 
in the signal region deduced from the fit ($N_{\rm b}$), which is 
consistent with the expectation from the background MC simulation 
($N_{\rm sig}^{\rm MC}$). 

The branching fractions are calculated as
${\cal B} = N_{s}/(2\cdot\varepsilon\cdot N_{B^{+}B^{-}})$
where $N_{B^{+}B^{-}}$ is the number of $\Upsilon(4S)\rightarrow B^{+}B^{-}$ 
events, assumed to be half of the number of produced $B$  meson pairs. 
The efficiency is defined as 
$\varepsilon = \varepsilon^{\rm tag}\times\varepsilon^{\rm sel}$,
where $\varepsilon^{\rm tag}$ is the tag reconstruction efficiency for events with 
$B^{-}\rightarrow\tau^{-}\bar{\nu}_{\tau}$ decays on the signal side, 
%determined by MC to be $0.136\pm 0.001({\rm stat})\%$,  
and $\varepsilon^{\rm sel}$ is the event selection 
efficiency listed in Table~\ref{tab:signal_yields}.
%, as 
%determined by the ratio of the number of events surviving all of the 
%selection criteria including the $\tau$ decay branching fractions over the number of fully 
%reconstructed $B^{\pm}$.
When all $\tau$ decay modes are combined
we obtain a branching fraction of $(1.06^{+0.34}_{-0.28}) \times 10^{-4}$.
The branching fraction for each $\tau$ decay mode is consistent within  
error as shown in Table~\ref{tab:signal_yields}.

\begin{table*}
  \renewcommand{\baselinestretch}{1.3}
    \begin{tabular}{ccccccccc} \hline
&$N_{\rm side}^{\rm obs}$ &$N_{\rm side}^{\rm MC}$ &$N_{\rm sig}^{\rm MC}$ &$N_{\rm obs}$ &$N_{\rm s}$ &$N_{\rm b}$
&$\varepsilon^{\rm sel}(\%)$  &${\cal B}(10^{-4})$  \\\hline %&$\Sigma$\\\hline
$\mu^{-}\bar{\nu}_{\mu}\nu_{\tau}$ &$96$   &$94.2\pm 8.0$   &$9.4\pm 2.6$  &$13$ &$5.4^{+3.2}_{-2.2}$  &$9.1^{+0.2}_{-0.1}$ &$8.88\pm 0.05$   &$1.01^{+0.59}_{-0.41}$ \\%&$2.0\sigma$\\
$e^{-}\bar{\nu}_{e}\nu_{\tau}$     &$93$   &$89.6\pm 8.0$   &$8.6\pm 2.3$  &$12$ &$3.9^{+3.5}_{-2.5}$  &$9.2^{+0.2}_{-0.2}$ &$8.18\pm 0.05$   &$0.79^{+0.71}_{-0.49}$ \\%&$1.3\sigma$\\
$\pi^{-}\nu_{\tau}$                &$43$   &$41.3\pm 6.2$   &$4.7\pm 1.7$  &$9$  &$3.4^{+2.6}_{-1.6}$  &$4.0^{+0.2}_{-0.1}$ &$5.79\pm 0.04$   &$0.96^{+0.74}_{-0.46}$ \\%&$1.9\sigma$\\
$\pi^{-}\pi^{0}\nu_{\tau}$         &$21$   &$23.3\pm 4.7$   &$5.9\pm 1.9$  &$11$ &$6.2^{+3.9}_{-2.7}$  &$4.2^{+0.3}_{-0.3}$ &$8.32\pm 0.08$   &$1.23^{+0.77}_{-0.53}$ \\%&$2.3\sigma$\\
$\pi^{-}\pi^{+}\pi^{-}\nu_{\tau}$  &$21$   &$18.5\pm 4.1$   &$4.2\pm 1.6$  &$9$  &$3.1^{+3.1}_{-2.6}$  &$3.7^{+0.3}_{-0.2}$ &$1.75\pm 0.03$   &$2.99^{+3.01}_{-2.49}$ \\%&$1.2\sigma$\\
\hline
Combined                           &$274$  &$266.9\pm 14.3$ &$32.8\pm 4.6$ &$54$ &$21.2^{+6.7}_{-5.7}$ &$30.2^{+0.5}_{-0.4}$ &$32.92\pm 0.12$  &$1.06^{+0.34}_{-0.28}$   \\%&$4.0\sigma$\\

\hline
    \end{tabular}
    \caption{The number of observed events in data in the sideband region $(N_{\rm side}^{\rm obs})$,
      number of background MC events in the sideband region $(N_{\rm side}^{\rm MC})$ and the 
      signal region $(N_{\rm sig}^{\rm MC})$,
      number of observed events in data in the signal region $(N_{\rm obs})$, 
      number of signal $(N_{\rm s})$ and background $(N_{\rm b})$ in the signal region determined by the fit, 
      signal selection efficiencies $(\varepsilon^{\rm sel})$, 
      extracted branching fraction $({\cal B})$ for $B^{-}\rightarrow\tau^{-}\bar{\nu}_{\tau}$.
      The listed errors are statistical 
      only. 
      %The last column gives the significance of the signal including 
      %the systematic uncertainty in the signal yield ($\Sigma$).
    }
   \label{tab:signal_yields}
\end{table*}

%==========================================================================
Systematic errors for the measured branching fraction are associated with 
the uncertainties in the  number of $B^{+}B^{-}$, signal yields and  
efficiencies.
The total fractional uncertainty of the combined measurement is 
$^{+20.5}_{-24.0}\%$, 
%$^{+16.9}_{-15.3}\%$, 
and we measure the branching fraction to be
$$
{\cal B}(B^{-}\rightarrow\tau^{-}\bar{\nu}_{\tau}) 
= (1.06^{+0.34}_{-0.28}(\mbox{stat})^{+0.22}_{-0.25}(\mbox{syst}))\times 10^{-4}.
%= (1.06^{+0.34}_{-0.28}(\mbox{stat})^{+0.18}_{-0.16}(\mbox{syst}))\times 10^{-4}.
$$
The significance is $4.0\sigma$ when all $\tau$ decay modes are combined, 
where the significance is defined as 
$\Sigma = \sqrt{-2\ln({\cal L}_{0}/{\cal L}_{\rm max})}$,
where ${\cal L}_{\rm max}$ and ${\cal L}_{0}$ denote the maximum likelihood 
value and likelihood value obtained assuming zero signal events, respectively.
Here the likelihood function from the fit is convolved with a Gaussian
systematic error function in order to include the systematic uncertainty
in the signal yield.

\section*{Acknowledgments}
The author wish to thank the KEKB accelerator group for the excellent operation of the KEKB
accelerator.

%\section*{Appendix}
% We can insert an appendix here and place equations so that they are
%given numbers such as Eq.~\ref{eq:app}.
%\be
%x = y.
%\label{eq:app}
%\ee

\end{document}